\documentclass{pasj00}
\draft
\usepackage{url}
\SetRunningHead{N. Sekiya et al.}{O\emissiontype{I} Fluorescent Line Contamination in SXDB Obtained with Suzaku/XIS}
\Received{}
\Accepted{}
\Published{}
\begin{document}
\title{O\emissiontype{I} Fluorescent Line Contamination in Soft X-Ray Diffuse Background Obtained with Suzaku/XIS}
\author{Norio \textsc{Sekiya}, Noriko Y. \textsc{Yamasaki}, Kazuhisa \textsc{Mitsuda}, and Yoh \textsc{Takei}}
\affil{Institute of Space and Astronautical Science, Japan Aerospace Exploration Agency, 3-1-1 Yoshinodai, Chuo, Sagamihara, Kanagawa 252-5210 Japan}
\email{sekiya@astro.isas.jaxa.jp}
\maketitle
\begin{abstract}
The quantitative measurement of O\emissiontype{VII} line intensity is a powerful method for understanding the soft X-ray diffuse background. 
By systematically analyzing the O\emissiontype{VII} line intensity in 145 high-latitude Suzaku/XIS observations, the flux of O\emissiontype{I} fluorescent line in the XIS spectrum, contaminating the O\emissiontype{VII} line, is found to have an increasing trend with time especially after 2011. 
For these observations, the O\emissiontype{VII} line intensity would be overestimated unless taking into consideration the O\emissiontype{I} fluorescent line contamination. 
Since the O\emissiontype{I} line emission originates from solar X-rays, this increase suggests that incident solar X-rays at the O\emissiontype{I} fluorescence energy tend to be larger than the early phase of Suzaku observations (2005 -- 2010). 
\end{abstract}

\section{Introduction}
The soft X-ray diffuse background (SXDB; \cite{1990ARA&A..28..657M}) is distributed over the whole sky.
In the SXDB below 1 keV, the blend of the O\emissiontype{VII} K$\alpha$ triplet line emission (centroids of the resonance, inter-combination and forbidden lines: 0.574, 0.569 and 0.561 keV, respectively) is one of the most prominent constituents \citep{2002ApJ...576..188M}. 
\citet{2009PASJ...61..805Y} suggests that it arises from two origins: approximately uniform emission of $\sim2$ photons s$^{-1}$ cm$^{-2}$ sr$^{-1}$ (LU), and spatially variable emission (0 -- 7 LU) from hot gas of temperature $kT\sim0.2$ keV.
The former is considered to originate from the heliospheric solar wind charge exchange (SWCX; \cite{1998LNP...506..121C, 2000ApJ...532L.153C, 2004A&A...418..143L}) plus the local hot bubble (LHB; \cite{1990ARA&A..28..657M}), and the latter from hot gas of the Galactic halo \citep{2000ApJ...543..195K}.
\citet{2011ApJ...726...91K} provides constraints on the heliospheric SWCX$+$LHB contributions to the local SXDB.
\citet{2013PASJ...65...32Y} gives an indication of long-term time variability of the O\emissiontype{VII} line intensity in the heliospheric SWCX due to the 11-year solar activity.
The physical state of the hot gas of the Galactic halo has been investigated with O\emissiontype{VII} absorption lines (e.g. \cite{2008ApJ...672L..21Y}; \cite{2010PASJ...62..723H}; \cite{2012ApJ...756L...8G}; \cite{2013ApJ...770..118M}).
The O\emissiontype{VII} line is used as a probe to understand about surrounding plasma around the solar system and the Galaxy.
For the study of the O\emissiontype{VII} line emission in the SXDB, X-ray Imaging Spectrometers (XISs; \cite{2007PASJ...59S..23K}) aboard Suzaku \citep{2007PASJ...59S...1M} are best suited because of their low and stable non X-ray background (NXB) and better energy response than those of CCDs onboard XMM-Newton or Chandra.
The O\emissiontype{VII} line can be contaminated by the O\emissiontype{I} K$\alpha$ line (centroid: 0.525 keV) because CCDs such as Suzaku/XIS cannot completely distinguish the O\emissiontype{I} line from the O\emissiontype{VII} line owing to their energy resolution of $\sim0.1$ keV (FWHM at $\sim1$ keV).
The O\emissiontype{I} line is created by fluorescence of solar X-rays with neutral oxygen in the Earth's atmosphere.
It is prescriptively thought to be removable by discarding time periods when the telescope is looking at the bright Earth or its atmosphere (e.g. elevation angle from bright Earth limb $>20^\circ$).
We analyzed 145 observations with the Suzaku/XIS from 2005 to 2012 and investigated the O\emissiontype{VII} line intensity in the SXDB with careful attention to the O\emissiontype{I} fluorescent line contamination.
In this letter, we report on an increasing trend of the O\emissiontype{I} fluorescent line intensity with time, especially after 2011, found during the analysis.
The error ranges stated in this letter show 90\% confidence level from the center value, and the vertical error bars in the figures indicate 1$\sigma$ level.
\section{Observations \& Data Reduction}
In order to investigate the O\emissiontype{VII} line intensity in the SXDB, we selected observations which have Galactic latitude $|b|>20^\circ$ to separate them from the X-ray emission peculiar to Galactic disk, and are apart from local diffuse emission structures such as super nova remnants and super bubbles (e.g. Loop I, Cygnus loop and so on).
In total, 145 observations were selected.
In this letter, we used cleaned event files of a back-illuminated CCD (XIS1) which has larger effective area than the sum of front-illuminated CCDs (XIS0, 2 and 3) below 1 keV.
We adopted the nominal data selection criteria following the standard pipeline selection criteria: elevation angle from bright/dark Earth limb $>$ $20^\circ$/$5^\circ$, and the Cut Off Rigidity (COR2) $>$ 8 GV $c^{-1}$ to reduce high energy particles background due to the low Earth's magnetic field.
For all the data sets, point sources whose flux were larger than $1.0\times10^{-14}$ erg cm$^{-2}$ s$^{-1}$ in the 0.5 -- 2.0 keV band were detected in the field of view and were removed with circular regions centered at their positions.
The radius of the circular regions were $>1.5'$ determined so that these regions included $>90$\% of source photons.
We also eliminated time periods in these observations when the proton flux in the solar wind exceeded the typical threshold, $4.0\times10^8$ cm$^{-2}$ s$^{-1}$ to reduce effects of the geocoronal SWCX (e.g. \cite{2009PASJ...61..805Y}).
These proton flux were observed with monitoring satellites; ACE/SWEPAM\footnote[1]{\url{http://www.srl.caltech.edu/ACE/ASC/level2/lvl2DATA_SWEPAM.html}} 
and WIND/SWE\footnote[2]{\url{http://web.mit.edu/space/www/wind_data.html}}.
The NXB was estimated from an accumulated night Earth observations with the ftool {\tt xisnxbgen} \citep{2008PASJ...60S..11T}.
For spectral analysis, the energy redistribution matrix files and ancillary response files were generated by the ftools {\tt xisrmfgen} and {\tt xissimarfgen} \citep{2007PASJ...59S.113I}, respectively.
In {\tt xissimarfgen}, we assumed a uniform sky with a radius of 20 arcmin.
HEAsoft version 6.12 and XSPEC version 12.8.0 were utilized.
\section{Analysis \& Results}
\begin{figure*}[htbp!]
  \begin{center}
    \includegraphics[width=\linewidth]{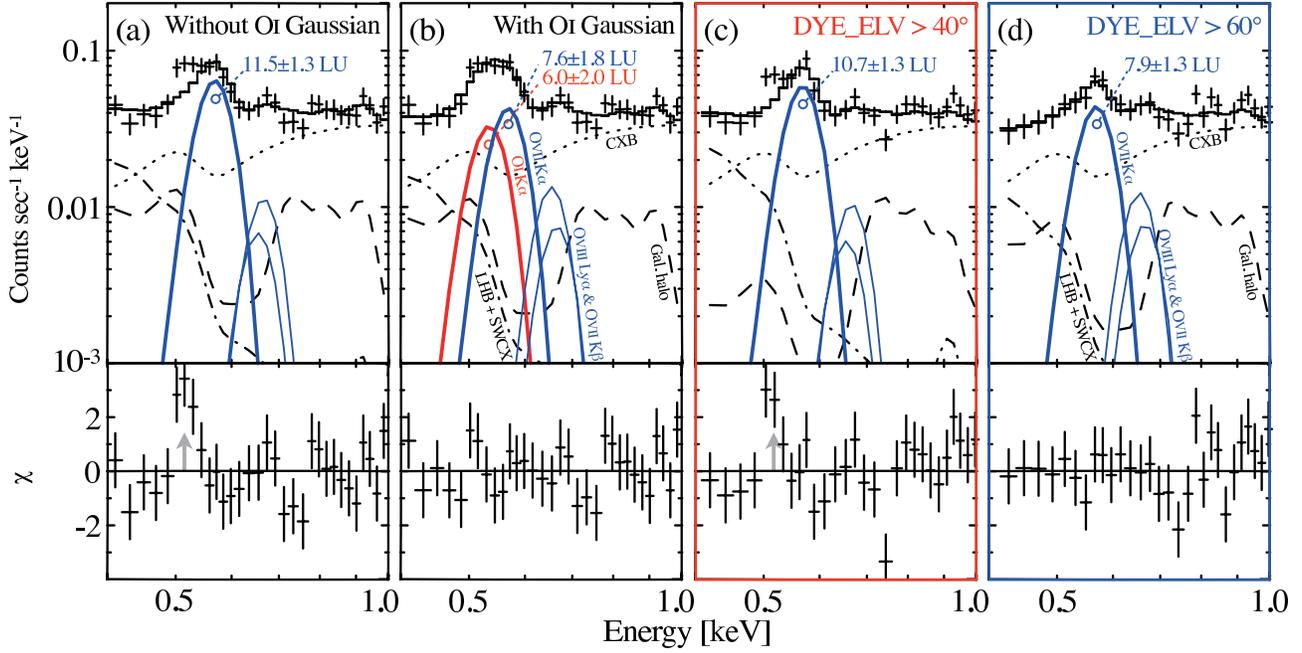}
  \end{center}
  \caption{0.4 -- 1.0 keV Suzaku/XIS1 spectra, best-fit emission models and residuals of LOCK\_365, without/with the O\emissiontype{I} Gaussian are shown in (b)/otherwise.
  The criteria of DYE\_ELV $>20^\circ$ (a and b), $>40^\circ$ (c) and $>60^\circ$ (d).}
  \label{fig:1}
\end{figure*}

\begin{figure}[htbp!]
  \begin{center}
    \includegraphics[width=\linewidth]{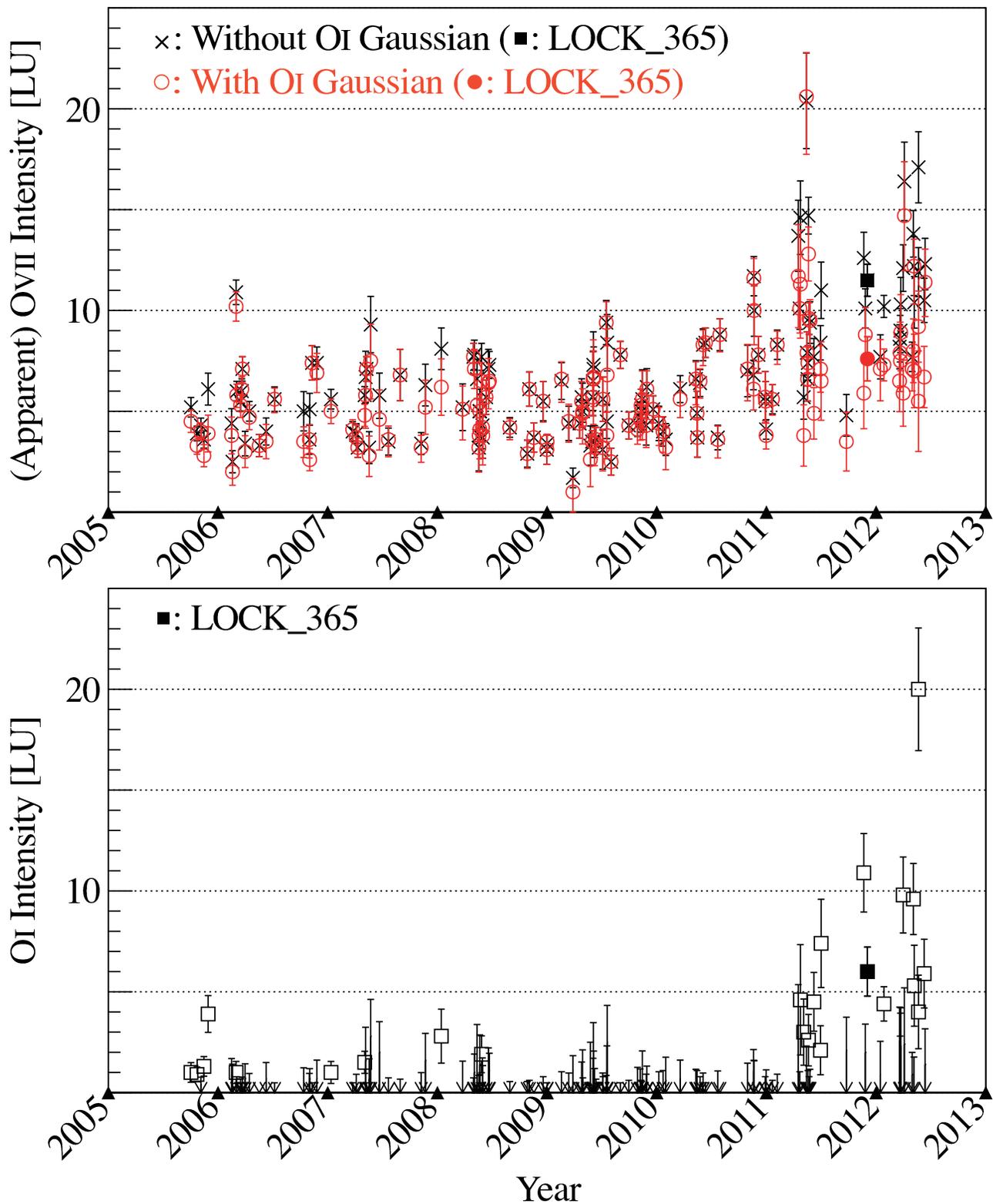}
  \end{center}
  \caption{Long-term time dependence of the O\emissiontype{VII}/O\emissiontype{I} intensity (top/bottom) in the 145 observations.
  Top: these O\emissiontype{VII} intensity are estimated without/with adding the O\emissiontype{I} Gaussian (black/red).
  Bottom: these O\emissiontype{I} intensity are estimated from the added O\emissiontype{I} Gaussian.
  }
  \label{fig:2}
\end{figure}

\begin{figure}[htbp!]
  \begin{center}
    \includegraphics[width=\linewidth]{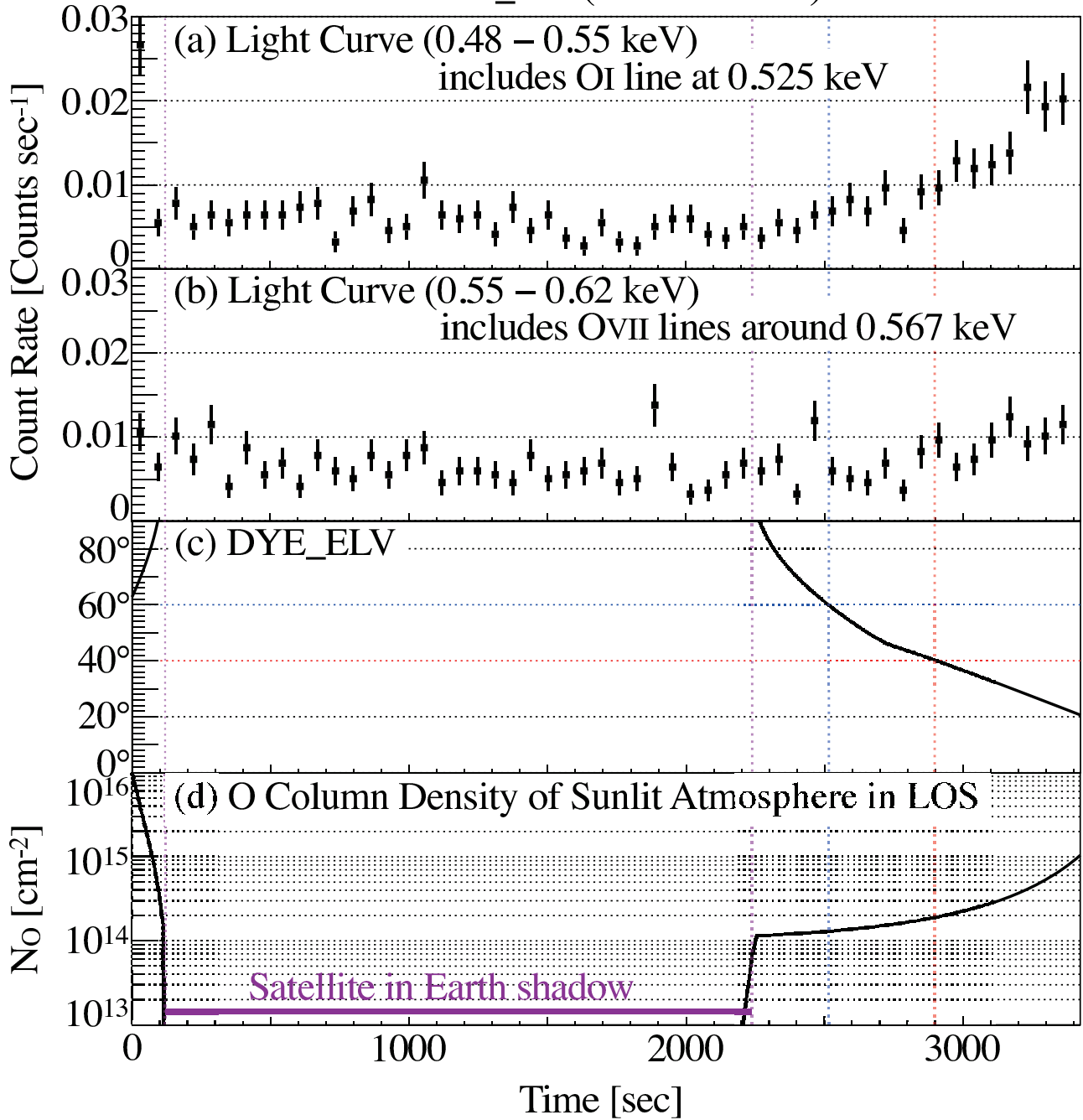}
  \end{center}
  \caption{Suzaku/XIS light curve for the O\emissiontype{I}/O\emissiontype{VII} folded by the orbital period are shown in (a)/(b). 
  The light curves are binned with 64 sec bin$^{-1}$.
  Temporal transition of DYE\_ELV and the O column density of sunlit atmosphere in the Suzaku line of sight using the MSIS model are shown in (c) and (d), respectively.
}
  \label{fig:3}
\end{figure}

\begin{figure}[htbp!]
  \begin{center}
    \includegraphics[width=\linewidth]{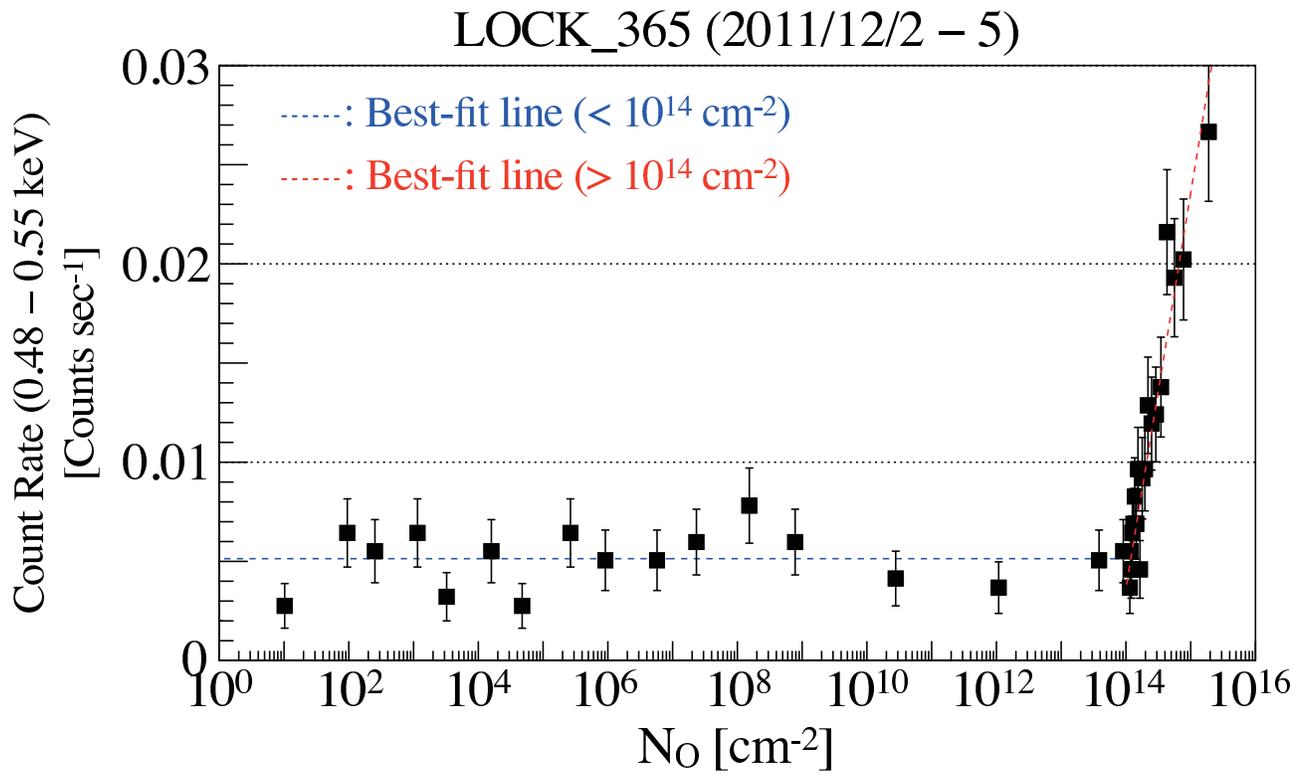}
  \end{center}
  \caption{The relation between the O column density of sunlit atmosphere in the Suzaku line of sight using the MSIS model and the 0.48 -- 0.55 keV count rate including the O\emissiontype{I}.
}
  \label{fig:4}
\end{figure}

We performed spectral analysis for all the Suzaku/XIS observations, according to the method shown in \citet{2009PASJ...61..805Y} and \citet{2013PASJ...65...32Y} in detail.
We first fitted the spectra in energy band of 0.4 -- 5.0 keV with the typical SXDB emission model consisting of (1) an unabsorbed optically-thin thermal collisionally-ionized (CIE) plasma, (2) an absorbed optically-thin thermal CIE plasma and (3) unresolved extragalactic point sources (cosmic X-ray background; CXB).
The first two components represent emission from heliospheric SWCX plus LHB blend and the Galactic halo.
As the optically-thin thermal CIE emission model, we used APEC (version 2.0.1; \cite{2001ApJ...556L..91S}; \cite{2012ApJ...756..128F}).
The abundance and redshift for both two APEC models were set to 1 solar \citep{1989GeCoA..53..197A} and zero, respectively. 
The temperature of the former APEC was fixed to 0.099 keV \citep{2013PASJ...65...32Y}.
The CXB is represented by absorbed double broken power-law with photon indices of 1.54 and 1.96 below 1.2 keV and a photon index of 1.4 above the energy \citep{2007PASJ...59S.141S}.
We fixed the normalization of the broken power law component with a low-energy photon index of 1.54 to 5.7 photons s$^{-1}$ keV$^{-1}$ sr$^{-1}$ at 1 keV.
The following model was adopted in the XSPEC software: apec$_{\rm SWCX+LHB}$ + phabs$_{\rm Gal.}$ $\times$ (apec$_{\rm Gal. halo}$ + 2$\times$bknpower$_{\rm CXB}$) where phabs$_{\rm Gal.}$ was for a photoelectric absorption in neutral hydrogen column densities ($N_{\rm H}$) of our Galaxy (LAB survey; \cite{2005AA...440..775K}).
Then the O\emissiontype{VII} (and O\emissiontype{VIII}) line intensity were estimated by adding three Gaussian lines (width = 0) at the O\emissiontype{VII} K$\alpha$, O\emissiontype{VlII} Ly$\alpha$, and O\emissiontype{VII} K$\beta$ energy (0.567 keV: averaging the centroids of resonance and forbidden lines, 0.653 keV and 0.666 keV, respectively), and modifying the original model by fixing the Galactic halo temperature at the best fit values, and setting oxygen abundance to zero for both the heliospheric SWCX$+$LHB and the Galactic halo.
When we use the APEC model with zero oxygen abundance, the contribution of oxygen radiative recombination continuum also disappears as an untrue side effect.
It was confirmed that this effect made only a few \% impact on estimation of the O\emissiontype{VII} line intensity \citep{2013PASJ...65...32Y}.
The fraction of the O\emissiontype{VII} K$\beta$ to the O\emissiontype{VII} K$\alpha$ was also fixed at 8.3\% \citep{2003ApJ...585L..73K, 2009PASJ...61..805Y}.
Hereafter, ``O\emissiontype{VII}'' is used to refer to the blend of the O\emissiontype{VII} K$\alpha$ line.
In some spectral fit results, however, residuals remain on the low energy side of the O\emissiontype{VII}, especially after 2011.
As an example, spectrum of ``LOCK\_365'' (Obs. ID: 806077010, Obs. date: 2011 Dec. 2 -- 5) is presented in Fig. \ref{fig:1} (a).
It is 0.7$^\circ$ distant from the Lockman hole where Suzaku observed annually for calibration purpose and was studied by \citet{2013PASJ...65...32Y}.
The apparent O\emissiontype{VII} intensity is estimated to be $11.5\pm1.3$ LU ($\chi^2$/d.o.f $=138.9/101=1.38$).
If energy and width of the O\emissiontype{VII} Gaussian are free, fitting result of energy, width ($\sigma$) and the apparent O\emissiontype{VII} intensity are $0.554\pm0.004$ keV, $0.019\pm0.007$ keV and $13.6^{+1.6}_{-1.5}$ LU, respectively ($\chi^2$/d.o.f $=114.8/99=1.16$).
The line centroid energy is inconsistent with the O\emissiontype{VII} energy, indicating a lower-energy line such as O\emissiontype{I}.
In this letter, the O\emissiontype{I} excluded the contribution of O$_2$ molecules but only included that of O atoms, because the former was negligible in any Suzaku observations whose line of sight did not penetrate the Earth's low atmosphere.
As shown in Fig. \ref{fig:1} (b), when we added a Gaussian line at the O\emissiontype{I} K$\alpha$ energy ($E_{\rm O}=0.525$ keV), the spectrum was better represented by the model and the goodness of the fit improved ($\chi^2$/d.o.f $=114.7/100=1.15$).
The O\emissiontype{VII} and O\emissiontype{I} intensity become $7.6\pm1.8$ LU and $6.0\pm2.0$ LU, respectively. 
If energy of the O\emissiontype{I} Gaussian is free, resultant energy, $0.527^{+0.015}_{-0.014}$ keV is consistent with $E_{\rm O}$.
Fig. \ref{fig:2} shows time dependence of the O\emissiontype{VII} (top) and O\emissiontype{I} (bottom) intensity in the 145 observations from 2005 to 2012, indicating significant increasing tendency of the O\emissiontype{I} intensity after 2011.
The O\emissiontype{VII} intensity of approximately half of observations after 2011 is overestimated without including the O\emissiontype{I} Gaussian.
We constructed the light curve of the LOCK\_365 observation folded by the orbital period, and checked the relation between the count rate and the elevation angle from bright Earth limb (hereafter DYE\_ELV) as shown in Fig. \ref{fig:3}.
Fig. \ref{fig:3} (a) shows 0.48 -- 0.55 keV band including the O\emissiontype{I} while Fig. \ref{fig:3} (b) shows 0.55 -- 0.62 keV band including the O\emissiontype{VII}.
Temporal transition of DYE\_ELV is shown in Fig. \ref{fig:3} (c). 
The count rate for the O\emissiontype{I} increases with decreasing DYE\_ELV while that for the O\emissiontype{VII}
varies little with DYE\_ELV.
The slight increase of the O\emissiontype{VII} count rate is due to the leakage of the O\emissiontype{I} in the 0.55 -- 0.62 keV band.
We retried to perform spectral analysis of LOCK\_365 with changing the criteria of DYE\_ELV from $20^\circ$ to $40^\circ$ and $60^\circ$ without the O\emissiontype{I} Gaussian as shown in Fig. \ref{fig:1} (c) and (d).
At the criteria of DYE\_ELV $=20^\circ$ and $40^\circ$, residuals exist on the low energy side of the O\emissiontype{VII}.
At the criteria of DYE\_ELV $=60^\circ$, on the other hand, the goodness of the fit improves ($\chi^2$/d.o.f $=95.0/96=0.99$) and the O\emissiontype{VII} intensity decreases to $7.9\pm1.3$ LU which is comparable to that estimated with the Gaussian representing O\emissiontype{I}.
Following previous studies (\cite{2007PASJ...59S.133F}; \cite{2007PASJ...59S.141S}; \cite{2008PASJ...60S..95M}; \cite{2009PASJ...61..805Y} and \cite{2013PASJ...65...32Y}), we calculated the O column density of sunlit atmosphere in the Suzaku line of sight using ``the MSISE-00 Model 2001" atmosphere model (hereafter MSIS model; \cite{1991JGR....96.1159H}; \cite{2002JGRA..107.1468P}), as shown in Fig. \ref{fig:3} (d).
Fig. \ref{fig:4} shows the relation between that column density and the 0.48 -- 0.55 keV count rate for the O\emissiontype{I} (the threshold of DYE\_ELV $=20^\circ$).
In the past analysis in \citet{2008PASJ...60S..95M} and \citet{2009Dthesis...Y}, they reported the O\emissiontype{I} emission only when the O column density is $>10^{15}$ cm$^{-2}$.
We reanalyzed the same data and confirmed their results.
In some observations after 2011 (e.g. LOCK\_365 observation in 2011), we detected clear O\emissiontype{I} increase from $10^{14}$ cm$^{-2}$ as shown in Fig. \ref{fig:4}.
However, in the other observations after 2011, no clear increase in O\emissiontype{I} band count rate is seen when the O column density $>10^{14}$ cm$^{-2}$ and only upper limits are obtained in O\emissiontype{I} intensity estimation.
\section{Discussion}
The O\emissiontype{I} line increased as significantly as detectable in Suzaku/XIS observations of the SXDB after 2011.
The origin of it is considered to be the Earth's atmosphere because of dependence on DYE\_ELV and correlation between the O column density and the 0.48 -- 0.55 keV count rate.
The best way to identify possible contamination and to remove the O\emissiontype{I} emission is to fold the light curve by the orbital period, or to sort by the O column density.
Adding a Gaussian for the O\emissiontype{I} in the spectral fitting may help exclude possible O\emissiontype{I} contamination when evaluating the O\emissiontype{VII} intensity.
The O\emissiontype{I} intensity ($I_{\rm O\emissiontype{I}}$) is following equation:
\begin{equation}
I_{\rm O\emissiontype{I}}=\frac{1}{4\pi}f_{\rm O\emissiontype{I}}N_{\rm O}\int F_{\rm Sun}(E)\sigma_{\rm O\emissiontype{I}}(E)dE
\end{equation}
where $f_{\rm O\emissiontype{I}}$, $N_{\rm O}$, $F_{\rm Sun}(E)$ and $\sigma_{\rm O\emissiontype{I}}(E)$ are the fluorescence yield of O\emissiontype{I} (0.0058; \cite{1979JPCRD...8..307K}), the O column density, the flux of the Sun at the Earth distance specified as a function of photon energy and the cross section for inner-shell ionization of O, respectively.
$F_{\rm Sun}(E)$ is derived from \citet{2001ApJ...560..499O} and $\sigma_{\rm O\emissiontype{I}}(E)$ is described in the Evaluated Nuclear Data File\footnote[3]{\url{http://t2.lanl.gov/nis/data/endf/endfvii-atomic.html}}.
Solar X-rays are mainly emitted from quiet regions and active regions of the solar corona.
The quiet regions produce nearly steady activities through the solar minimum and maximum. 
Considering the contribution of the quiet regions, the energy integral of $F_{\rm Sun}(E)\sigma_{\rm O\emissiontype{I}}(E)$ is about $2\times10^{-12}$ photons s$^{-1}$, and $I_{\rm O\emissiontype{I}}$ is
\begin{equation}
I_{\rm O\emissiontype{I}}\sim10^{-15}\ {\rm photons\ s^{-1}\ sr^{-1}}\times N_{\rm O}.
\end{equation}
Thus, $N_{\rm O}=10^{15}$ cm$^{-2}$ implies $I_{\rm O\emissiontype{I}}\sim1$ LU.
This calculation result supports the fact that the O\emissiontype{I} fluorescent line is seen when $N_{\rm O}>10^{15}$ cm$^{-2}$ in the early phase of Suzaku observations (2005 -- 2010); under the solar minimum.
On the other hand, the active regions of the solar corona significantly appear during the solar maximum.
The flux from the active regions is more than tenfold larger than that from the quiet regions.
It is compatible with the fact that the O\emissiontype{I} clearly appears when $N_{\rm O}>10^{14}$ cm$^{-2}$ in some observations after 2011 (e.g. LOCK 365); under the solar maximum.
In the other observations after 2011, no clear the O\emissiontype{I} fluorescent line is seen even when $N_{\rm O}>10^{14}$ cm$^{-2}$.
The reason can be that position and property of the active regions of the solar corona differ day by day.
Based on the above, a possible explanations for increasing tendency of the O\emissiontype{I} fluorescent line is conceivable; incident solar X-rays tend to be larger than the early phase of Suzaku observations (2005 -- 2010).

\bigskip
We are grateful to Dr. Randall K. Smith, the referee, for fruitful suggestions.
We acknowledge Kazuhiro Sakai for helpful comments about the analysis.
We also thank to the Suzaku, ACE and WIND team for providing the data.
This work is partly supported by a Grants-in-Aid for Scientific Research from JSPS (Project Number: 12J10673).
\bibliographystyle{bibstyle}
\bibliography{Sekiya_OI_contamination}


\end{document}